\documentclass[letterpaper,twocolumn,10pt]{article}
\usepackage{usenix,epsfig,endnotes}
\usepackage{cite}
\usepackage{amsmath,amssymb,amsfonts}
\usepackage{algorithmic}
\usepackage{graphicx}
\usepackage{textcomp}
\usepackage{xcolor}
\usepackage{subfigure}
\usepackage{pgfplots}
\usepackage{caption}
\usepackage{url}
\begin{document}

\date{}

\title{\Large \bf SeqNet: An Efficient Neural Network for Automatic Malware Detection }

\author{
{\rm Jiawei Xu}\\
Nankai University
\and
{\rm Wenxuan Fu$^*$}\\
Nankai University
\and
{\rm Haoyu Bu\thanks{Equal contribution}}\\
Nankai University
\and
{\rm Zhi Wang\thanks{Corresponding author}}\\
Nankai University
\and
{\rm Lingyun Ying}\\
QiAnxin Corporation
}
\maketitle

\thispagestyle{empty}

\subsection*{Abstract}
Malware continues to evolve rapidly, and more than 450,000 new samples are captured every day, which makes manual malware analysis impractical.
However, existing deep learning detection models need manual feature engineering or require high computational overhead for long training processes, which might be laborious to select feature space and difficult to retrain for mitigating model aging.
Therefore, a crucial requirement for a detector is to realize automatic and efficient detection.

In this paper, we propose a lightweight malware detection model called SeqNet which could be trained at high speed with low memory required on the raw binaries.
By avoiding contextual confusion and reducing semantic loss, SeqNet maintains the detection accuracy when reducing the number of parameters to only 136K.
We demonstrate the effectiveness of our methods and the low training cost requirement of SeqNet in our experiments.
Besides, we make our datasets and codes public to stimulate further academic research.

\section{Introduction}

Malware is a serious cyber security threat, which could cause severe damage to individual and corporate systems, for example, the dramatic slowdown or breakdown, critical data loss or leakage, and catastrophic hardware failure. 
AVTest reports that, on average, over 450,000 new malicious programs and potentially unwanted applications are detected every day.~\cite{avtest}.
The enormous volume of new malware variants renders manual malware analysis inefficient and time-consuming.




To detect malware more efficiently, many researchers proposed advanced tools for malware analysis and detection~\cite{deepreflect, Yarix, BinaryNinja, yara}.
These tools help analysts complete their tasks more efficiently by performing partial work on their behalves.
However, these solutions could not fundamentally reduce their workload when processing such a large number of malware.
To tackle this problem, many experts and scholars apply machine learning algorithms, especially deep learning, to malware detection and classification~\cite{2017Android, mtnet, Droid-Sec, deepflow, dynamic, MobiTive, SHLMD, Multimodal_with_various_android_features, VariousMethodsSynthesis, 2021Detecting, 2018Classification, deepmal, IMCEC, TrafficDetection_anotherDeepmal, DL-FHMC, RMVC, deeprefiner, robust-android, sigl, RobustPDF}.
Their efforts contribute a lot to the research of malware analysis neural networks and practical automatic malware detection.

However, these models usually require various feature engineering to help neural networks make judgments, which might be laborious and easy to lose some critical information.
To realize more user-friendly and automatic detection, binary-based methods have been proposed~\cite{malconv2, malconv, M2018Deep, deepmal}.
The two epidemic approaches for raw binary processing are file cutting and binary-image converting.
Nevertheless, these two methods probably suffer from contextual confusion and semantic loss, which will be discussed later.

Additionally, model aging is a crucial problem for neural networks~\cite{concept, space_time}.
Different from computer vision and natural language processing, malware is evolving continuously and rapidly.
Malware detection is a battle between attackers and detectors.
As malware continues to evolve, deep learning models might be out-of-date.
For example, the model trained five years ago might be very weak in malware detection today.
Neural networks are hard to recognize unseen malicious behaviors, which could cause lower detection accuracy and easier evasion.

It is impossible for models to predict the features of the future malware but feasible to quickly learn the knowledge of detecting new malware.
Therefore, retraining models becomes one of the few methods to mitigate the aging problem.
We can make neural networks quickly retrained and learn new features of new malware so that they could recognize novel attack approaches.

Due to the structure and the scale of the models, retraining might be time-and-computation consuming, and such a high cost might make model updating difficult.
Furthermore, malware detection is a usual operation in almost every electronic system, and it is necessary for devices with a low computational capacity to perform detection.
For example, it is difficult for laptops or other mobile devices to run a huge model to scan all the files for malware detection.
These requirements illustrate that the detection model should be small and efficient enough to make it more practical so that we could quickly retrain or perform detection.
Also, automatic detection without complex feature engineering is critical for models to be used in various scenarios.

Generally, there are two challenges in our work, automatic and efficient.
The detection model should be automatic enough, and it requires little manual feature engineering.
The scale of the model should be small enough so that it could have lower training and detection cost.

In this paper, we propose an efficient automatic malware detection model with only about 136K parameters and refer to it as SeqNet.
Without artificial feature selection, SeqNet could automatically analyze samples and find the differences between malicious and benign programs only based on raw binaries.

The smaller neural networks usually have fewer parameters, which might lead to lower learning capacity.
It might be because smaller models are often more challenging to fit the complex mapping from the raw binaries to the malicious possibility domain.
This problem probably causes lower malware detection accuracy in small deep learning models.

To maintain accuracy when reducing the number of parameters, we propose a novel binary code representation method to reduce semantic loss and avoid contextual confusion.
Depending on our method, we make SeqNet perform well on Portable Executable~(PE) malware detection without feature engineering.
Based on this representation approach, we create a new convolution approach, called Sequence Depthwise Separable Convolution~(SDSC), to further squeeze the scale of the detection model.

We train SeqNet on a large PE dataset and find it has great performance, compared to many existing binary-based methods and models.
We also demonstrate the effectiveness of our model shrinkage methods in further experiments.
Besides, we make our code and dataset public for further research, and we hope that deep learning algorithms will be applied to malware detection better.

The main contributions of this paper include:
\begin{enumerate}
    \item We propose a novel approach to representing binary code while reducing semantic loss and avoiding contextual confusion.
    \item Based on the new representation method above, we propose SDSC, a novel convolution method for squeezing malware detection models.
    \item We devise a deep Convolutional Neural Network~(CNN), called SeqNet, which has a much smaller size and shorter training process.
    \item We make our dataset and codes public for further research.
\end{enumerate}

Here is the layout of this paper.
Section~\ref{background} introduces dominant approaches and several problems in deep malware detection.
Section~\ref{methods} describes the major methodology we apply to SeqNet.
Section~\ref{experiments} elaborates our experiments and the corresponding results.


\section{Background}\label{background}
In this section, we introduce the background of malware detection using deep learning.
First we enumerate two main approaches in this area to our knowledge.
Then we discuss several problems of the popular binary representation methods further.
At last, we explain the depthwise separable convolution that one of our approaches is based on.
    
\subsection{Deep Malware Detection}
Neural networks have powerful learning abilities and have been widely used in computer vision and natural language processing.
Deep learning algorithms have already been applied to malware detection by many researchers.
To our knowledge, we consider that there are two mainstream ideas, which is similar to~\cite{explanation-attack}.

\noindent\textbf{Feature-based Methods.}
In early works, deep learning models are trained from carefully crafted malware features~\cite{MobiTive, DL-FHMC, SHLMD, Multimodal_with_various_android_features, 2018Classification, 2017Android, deeprefiner, robust-android, sigl}.
When checking a suspicious sample, models need to extract the specific features, process them in specific ways, and then detect malicious codes to give their results.
The selected features could be API calls, control flow graphs~(CFG), or any other information which is able to reflect the action of a program.
It is indeed that learning from manual features is an effective way for neural networks to recognize the main differences between malicious and benign samples.
However, the specific-domain features could only well characterize the crucial information of samples from one perspective.
It could not fully cover the binary code semantics and even triggers significant information loss.
For example, only using API calls as the feature would cause models to ignore the control flow.
Also, the crafted features require sufficient prior knowledge, which needs specialists to select carefully.
Therefore, time-consuming manual feature extraction probably limits the usage of the feature-based models and makes them difficult to combat the continuous evolution of malware.

\noindent\textbf{Binary-based Methods.}
Nowadays, automatic feature extraction is one of the trends of neural networks with less human intervention and better performance than traditional feature engineering.
We read the binary of a file and directly send it to the detection model without or with little preprocessing.
The model will automatically find the suspicious part and recognize the binary as malicious or benign.
This approach could more effectively avoid the need for people to analyze malware and better reduce the workload of analysts.
Also, learning directly from the raw binaries might be better in preserving semantic and contextual information by mitigating the loss caused by manual feature engineering.


To make our model more automatic and avoid information loss, we focus on the binary-based models, and SeqNet applies raw binaries as input.
Additionally, lower computational overhead could make models better adapt to the evolution of malware and expand the application scenarios, for example, in the IoT environment.
We apply a novel but simple binary code representation method to SeqNet and maintain its performance when reducing the parameters.
        

\subsection{Binary Code Representation}
\begin{figure*}[]
\centering
\subfigure[]{
\label{fig:confusion:edge}
\includegraphics[scale=0.4]{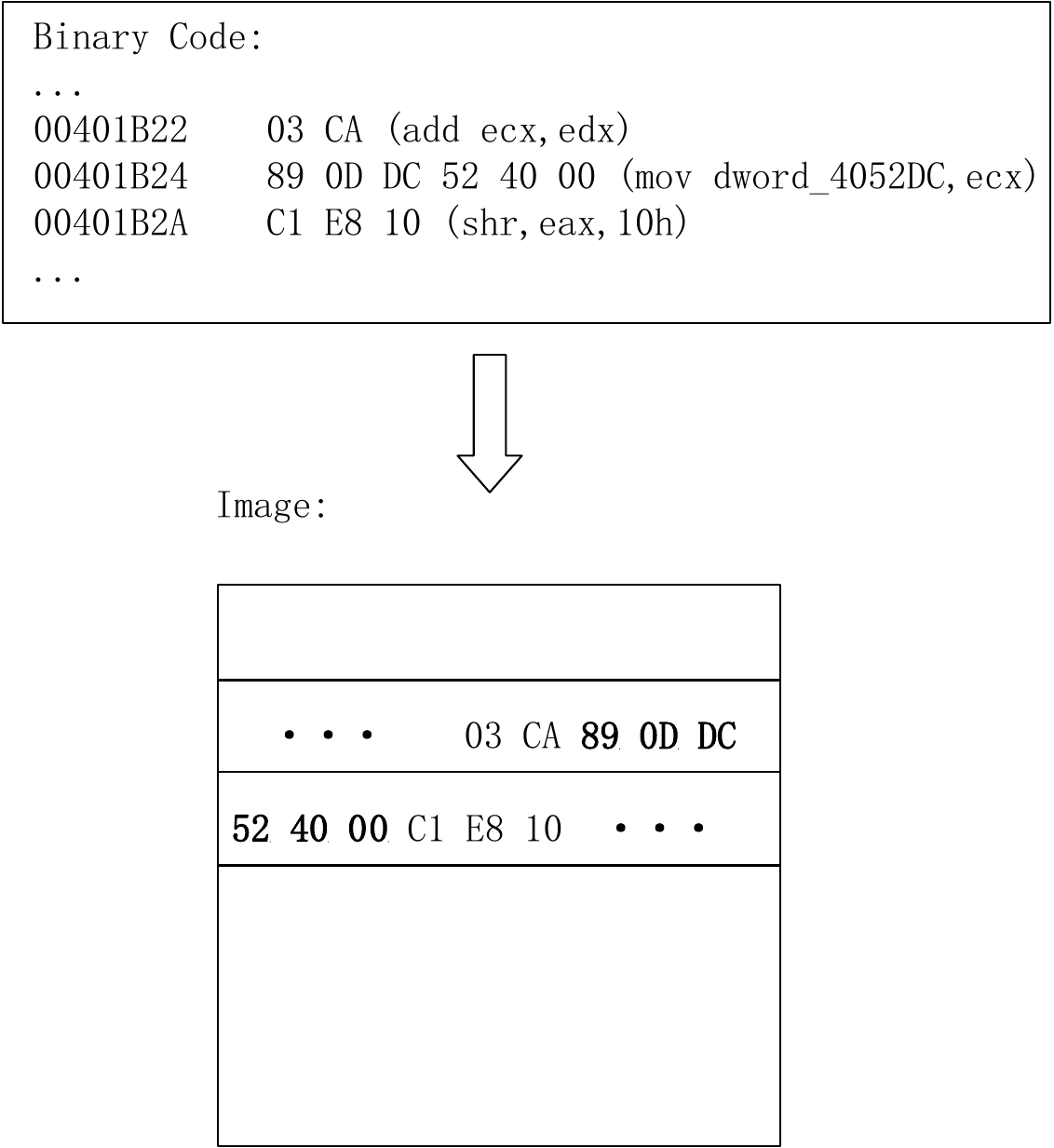}
}
\subfigure[]{
\hspace{2.5em}
\label{fig:confusion:context}
\includegraphics[scale=0.4]{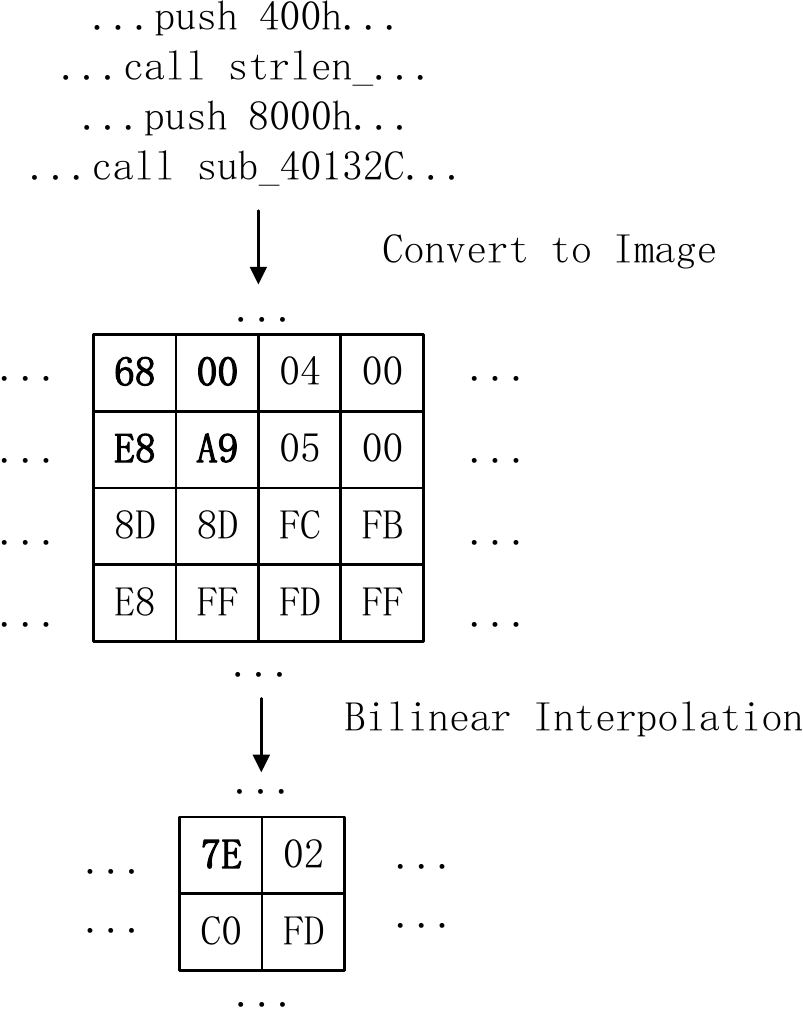}
}
\subfigure[]{
\hspace{2em}
\label{fig:confusion:padding}
\includegraphics[scale=0.4]{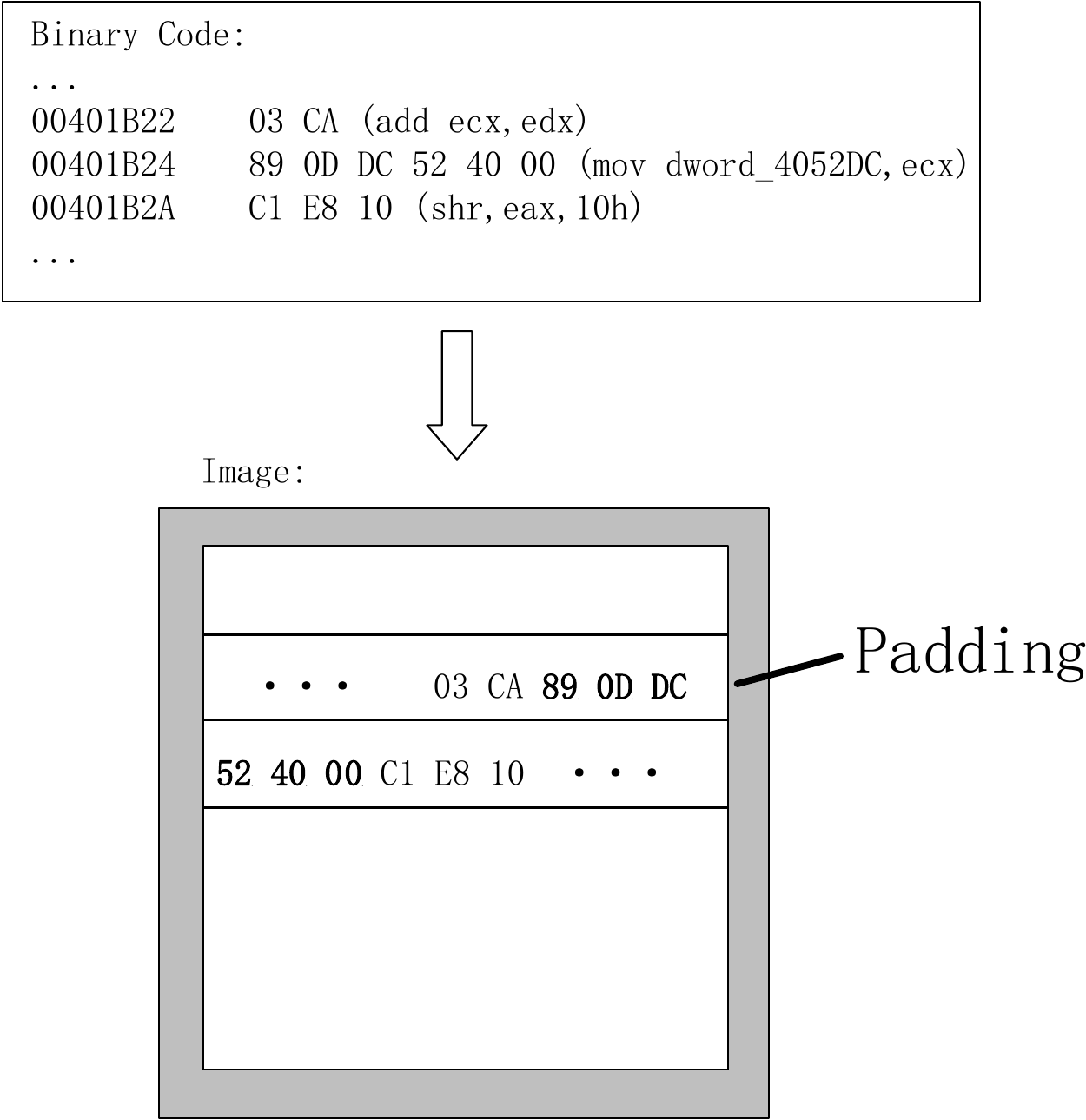}
}
\caption{
Examples of contextual confusion caused by binary-image converting. 
(a) Edge Loss: the instruction "89 0D DC 52 40 00" would be cut off at the edge of the image, and the convolution kernels could not well recognize this broken instruction, which would introduce edge loss. 
(b) Resampling Noise: In the last result, '7E' comes from '68 00' and 'E8 A9'. '68 00' represents the instruction 'push', and 'E8 A9' refers to 'call' instruction.
The bilinear interpolation algorithm imposes an improper contextual relationship between the two instructions, which would cause resampling noise.
(c) Padding Problem: The padding operation might mislead the model to believe the improper information at the edge of the image.
Besides, the instruction "89 0D DC 52 40 00" is disrupted further due to the padding.
}
\label{fig:confusion}
\end{figure*}

In this part, we will introduce several major binary code representation methods applied to binary-based models.
Translating samples into the input of neural networks could significantly affect the performance of the models.
Therefore, a proper binary code representation method is an important part of a binary-based malware detection neural network.
Nowadays, two main methods are proposed to fully represent the raw binary codes.

\noindent\textbf{File Cutting.}
Due to the limitation of memory constraints, many works set an artificial limit on the maximum file size, and this method is to take a fixed-length code snippet from the beginning of a binary program.
If the length of a binary program is less than the length of the required code snippet, the snippet would be padded with zeros at the end.
File cutting suffers from semantical information loss because the end of the binary will be ignored if it is much longer than the fixed length.
However, malicious codes are often located at the end of the binary files.
For example, embedded viruses usually embed themselves at the tail of the infected files, which might help them evade the detection of the models based on this approach.
To mitigate this problem, MalConvGCT~\cite{malconv2} improved MalConv~\cite{malconv} performance by expanding the snippet size limit.

\noindent\textbf{Binary-image Converting.}
The second method converts all binary codes into an image and leverages image classification solutions to perform malware detection. 
All images could be resampled to the same size using the bilinear interpolation algorithm.
However, images are different from sequences, which might result in several problems.
We consider that this method would cause contextual information confusion and the following lists three examples.

\begin{itemize}
    \item \textbf{Edge Loss:}
    If a binary instruction is located at the edge of the image, line breaks may cut off the instruction into two parts, as shown in Figure~\ref{fig:confusion:edge}. 
    This problem might cause the model hard to recognize several long instructions.
    Besides, contextual information is probably disrupted at the edge due to the break of strongly correlated instructions.
    
    \item \textbf{Resampling Noise:}
    If we reshape the image size, unrelated instructions in different lines introduce contextual information confusion, as shown in Figure~\ref{fig:confusion:context}.
    This problem easily makes the instructions far from each other in the original sequence forced to be integrated in the corresponding image, which might confuse the neural network.
    
    \item \textbf{Padding Problem:}
    The padding operation might make the model hard to recognize the beginning and the end of the original sequences, as shown in Figure~\ref{fig:confusion:padding}. 
    To ensure the consistency of the convolutional layer inputs and outputs, we usually pad some zeros at the edge of the input, and the neural network might get spatial information according to the padding~\cite{Translation_Invariance}. 
    Different from image processing, the recognized spatial information possibly misleads the model.
\end{itemize}

The semantic loss caused by file cutting and the contextual confusion caused by binary-image converting hinder the performance of malware detection models.
These issues might confuse neural networks and even mislead them to make diametrically opposed decisions.
By alleviating these problems, we help our model maintain its performance when reducing its parameters.
        

\subsection{Convolution Methods}
\begin{figure*}[h]
\centering
\subfigure[]{
\includegraphics[scale=0.5]{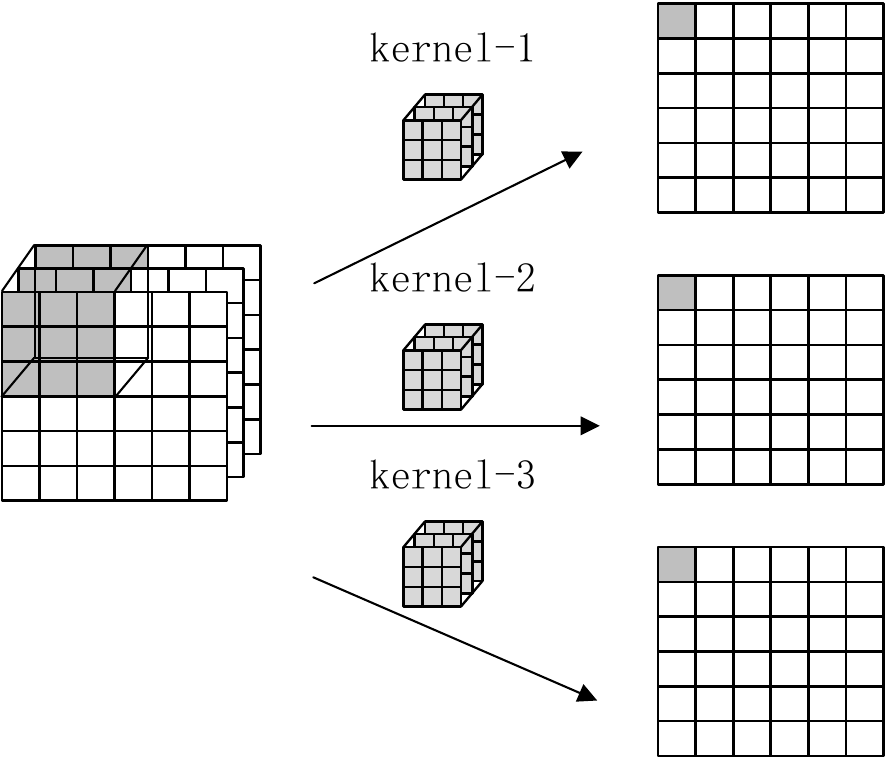}
\label{fig:depthwise:std}
}
\hspace{2em}
\subfigure[]{
\includegraphics[scale=0.5]{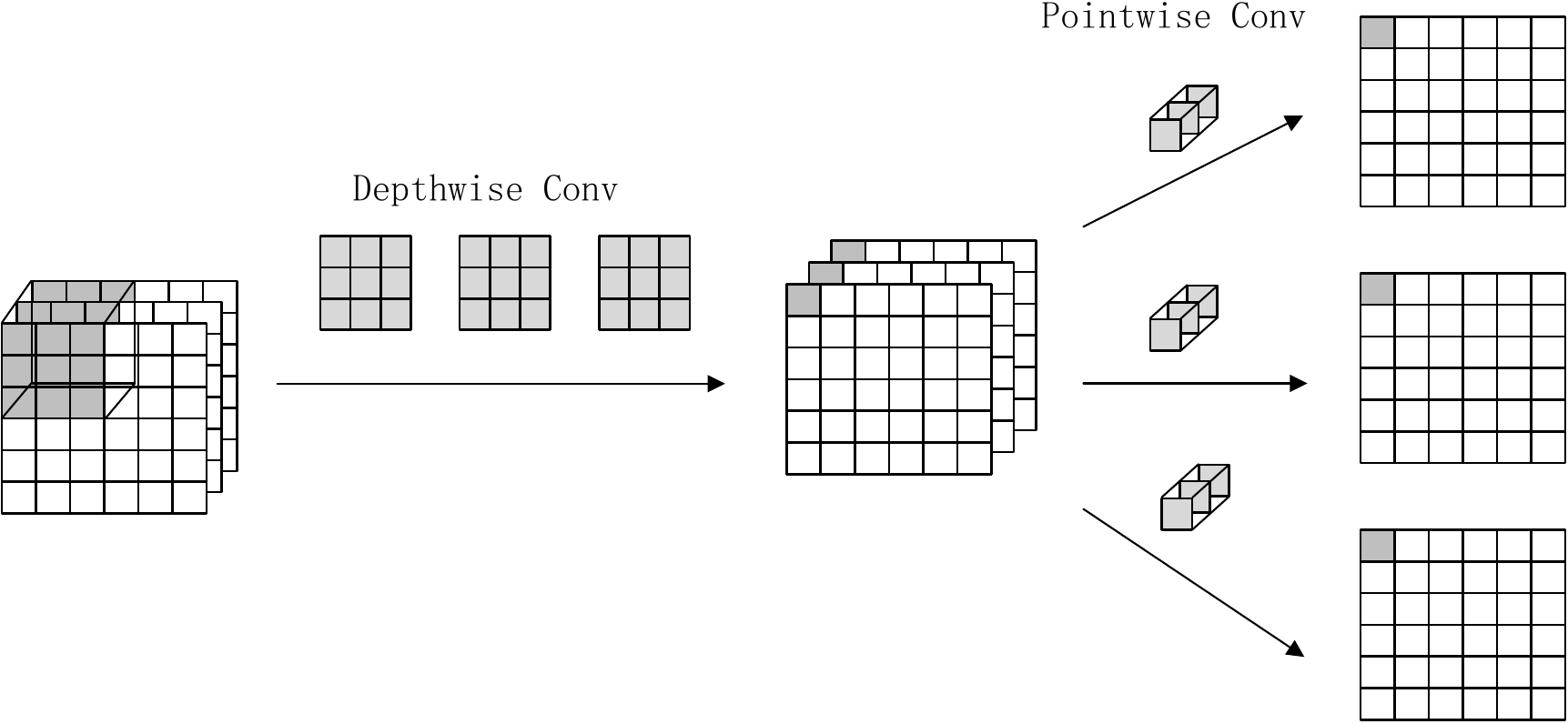}
\label{fig:depthwise:dep}
}
\subfigure[]{
\includegraphics[scale=0.8]{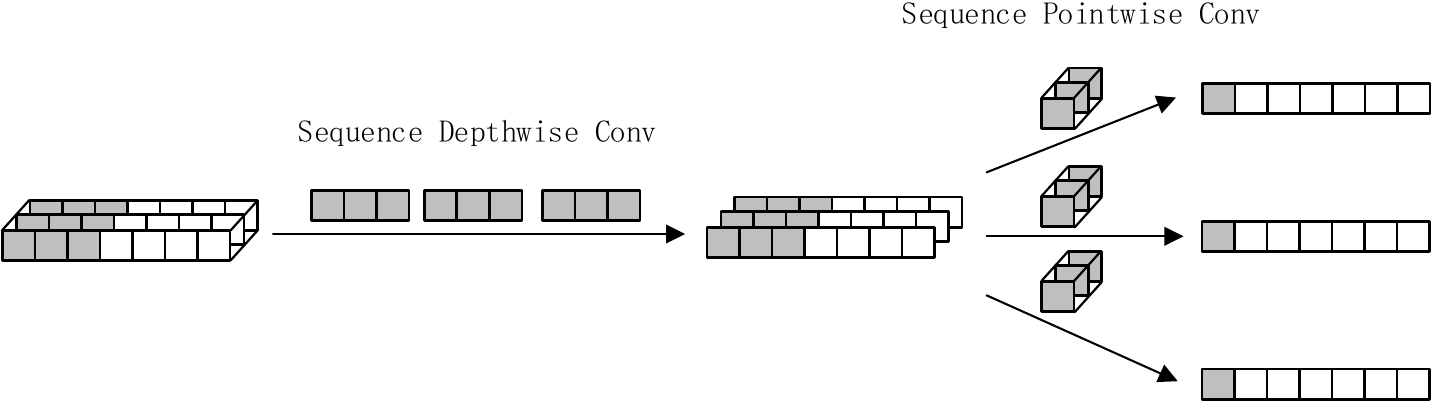}
\label{fig:depthwise:our}
}
\caption{Standard convolution shown in Figure~(a) uses one standard kernel to generate a corresponding channel.
Depthwise separable convolution shown in Figure~(b) factorizes the standard convolution into two parts and uses the pointwise kernels to create these channels with fewer parameters.
Sequence depthwise separable convolution shown in Figure~(c) changes the depthwise convolution kernels from 2D to 1D which further reduces the number of parameters.
}
\label{fig:depthwise}
\end{figure*}

Traditional convolution simulates animal vision and has great performance in computer vision.
This simple operation effectively extracts visual features in images.
Low-level convolutional layers detect the textural and simple features in images, and high-level convolutional layers could recognize the content and overall semantics~\cite{deconv}.
That is why computers can identify complex objects with the superposition of multiple convolutional layers.

However, the number of parameters needed in traditional convolution often makes deep learning models too large to apply on devices with low computational capacity.
Also, a model with too many parameters probably has a very long training process.
For example, VGG has more than 130 million parameters and has been trained for $2 \sim 3$ weeks~\cite{vgg}.
It is not suitable for such a huge model to run on common devices.

To squeeze the size of neural networks, advanced methods have been proposed, and depthwise separable convolution~(DSC) is one of them.
The DSC factorizes common convolution into depthwise convolution and pointwise convolution, which is proposed by Howard et al~\cite{mobilenet} and used in MobileNets.
As shown in Figure~\ref{fig:depthwise}, compared to standard convolution, DSC could effectively reduce the number of training parameters.
This method also makes MobileNets much smaller than many prior models and available to run on mobile devices.

Recent works have revealed that the CNNs could effectively find the salient statistical differences between samples in different classes~\cite{overinterpretation}.
Similar to the hypothesis in~\cite{deepreflect, vae-mlp-android}, we consider that malware might have special codes for attacks, and CNNs could explore the distance between malicious and benign samples.
Also, several researchers point out that DSC and self-attention mechanism~\cite{transformer} which is widely used in natural language processing tasks have similar effects on generating final results~\cite{conv-self-attention, attention-conv}.
Compared to the self-attention mechanism, DSC usually needs fewer parameters and lower computation overhead while processing long binaries.

Therefore, we apply CNN architecture to SeqNet and adapt the input form of DSC from 2D to 1D, which better reduces the number of training parameters. 
Accordingly, the training time cost and the size of the newly generated model are both reduced further.
The details of our approach are described in Section~\ref{methods}.

\section{Methods}\label{methods}
In this section, we will introduce the details of SeqNet.
Firstly, we give an overview of our approaches.
Secondly, we introduce the sequence characterization that could reduce semantic loss and avoid contextual confusion. 
The third part describes how SDSC compresses the scale of the model.
At last, we elaborate on the architecture of SeqNet.


\subsection{Overview}
The goal of SeqNet is to achieve efficient and automatic malware detection with low training costs.
During the whole training and detection processes, the operator does not need professional malware analysis knowledge to perform manual domain-specific feature engineering.
In practice, we directly input raw binaries into SeqNet, and SeqNet will automatically analyze the sequences and extract the features.
The output of SeqNet is the malicious possibility of a suspicious sample, and whether the input sample is malware or not is decided by the possibility given by the model.

Detecting accurately is an essential requirement of malware detection models.
We consider that malware detection is different from image classification. 
Malware detection might need more attention to several crucial malicious codes, while image classification might focus more on the whole.
Depending on this theory, one of our main design outlines of SeqNet is to reduce contextual confusion and semantic loss.
We use the raw binary sequence as the input of SeqNet, which can avoid contextual confusion and reduce semantic loss.

Lightweight models often have a wider range of application scenarios and faster detection performance.
It is obvious that small models also have low training costs.
Therefore, squeezing the scale of SeqNet is necessary. 
The new convolution method, called Sequence Depthwise Separable Convolution~(SDSC), helps SeqNet meet this requirement.


\subsection{Sequence Characterization}
\begin{figure*}
\centering
\includegraphics[scale=0.27]{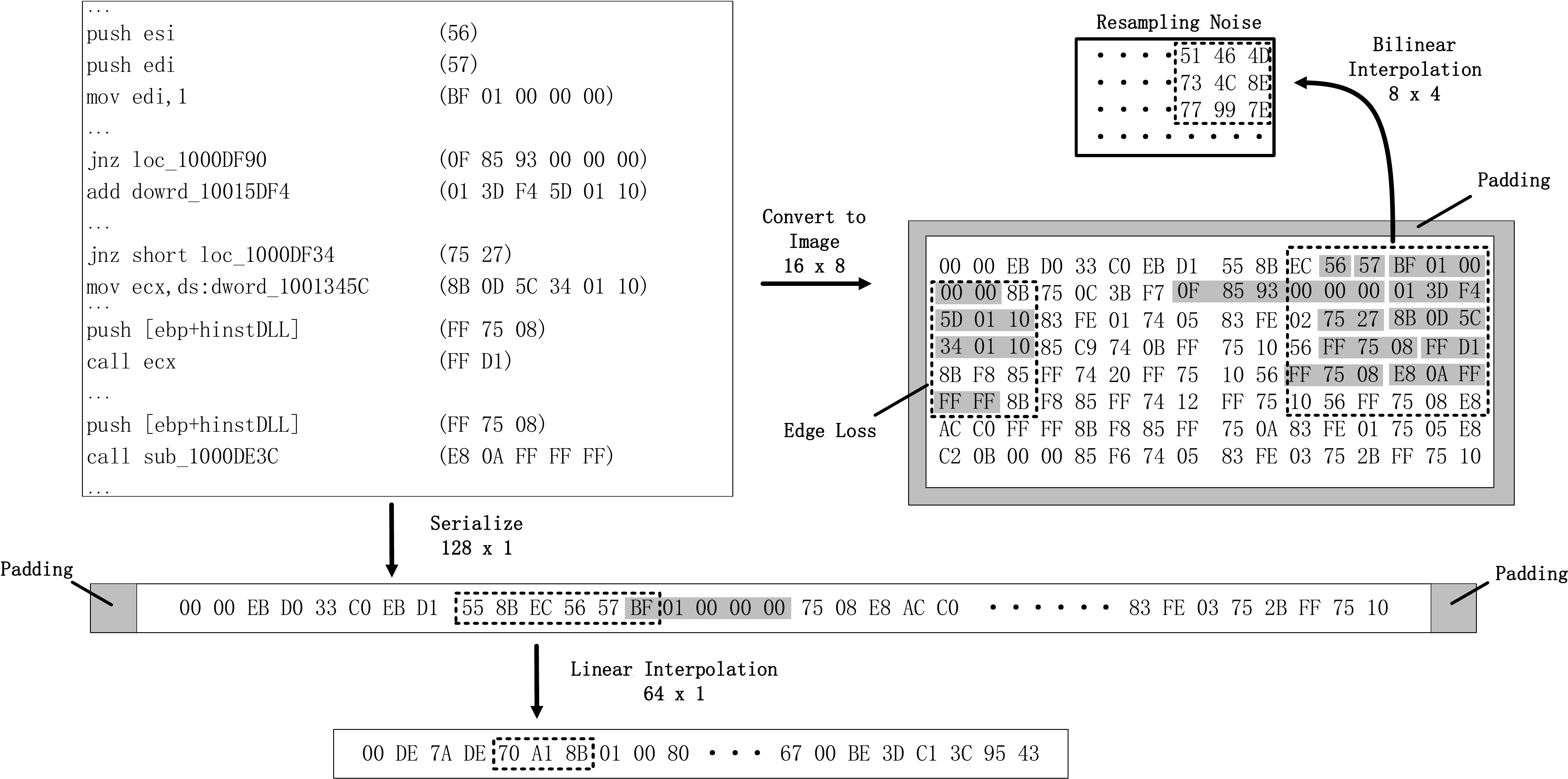}
\caption{
An example shows sequence characterization could address the problem of contextual confusion and semantic loss.
In the image, the binary instruction "BF 01 00 00 00" is cut off at the edge, but it remains the shape in the sequence.
After interpolating, we can see that the image forces to strengthen the relationship between "56" which means "push esi" and "00" in "0F 85 93 00 00 00" which means "jnz loc\_1000DF90", but it ignores the instructions with stronger relationship, for example, "push edi" and "mov edi, 1". 
In the sequence, the instruction "56" combines with the front instead of "00", and it remains physically close with "57" which means "push edi".
In the image, the padding which is added before inputting into convolutional layers provides improper location information, but in the sequence, it marks the beginning and the end.
}
\label{fig:solution}
\end{figure*}
\begin{figure}
\centering
\subfigure[]{
\includegraphics[scale=0.8]{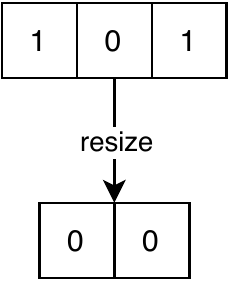}
\label{fig:scale:int}
}
\hspace{0.3em}
\subfigure[]{
\includegraphics[scale=0.8]{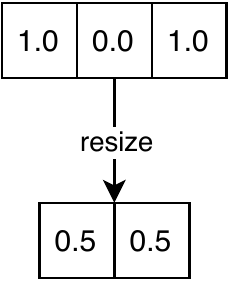}
\label{fig:scale:float}
}
\caption{
If we resize the sequence before normalizing, the result could not represent all information in the original binary codes~(a).
On the contrary, if we resize the sequence stored in float format after normalizing, we could reduce the semantic loss effectively~(b).
}
\label{fig:scale}
\end{figure}

The input format is critical for neural network performance and model size.
Larger input often leads to larger models, and proper input format could effectively improve the learning effect of neural networks.

The input of SeqNet is the raw binary sequences which are resized to the same length by the linear interpolation algorithm.
Raw binary sequence input needs little human intervention.
Without converting to images, it is obvious that we could avoid contextual information confusion and reduce semantic loss, as shown in Figure~\ref{fig:solution}.

\noindent \textbf{Edge Loss Avoidance.}
The edge loss could be avoided because the sequence only has two edges, the beginning and the end.
The sequence format conforms to the spatial structure of codes, so there is no break in any instruction, which makes all instructions intact when inputting into the model.
The disappearance of interruptions also effectively protects the semantics of the raw binary sequence because the instructions in close proximity are not separated.

\noindent \textbf{Resampling Noise Reduction.}
The resampling noise could be reduced because the elements could only be influenced by the forward and backward context when we resize the sequence.
Additionally, the imposed relationship in the image between two unrelated instructions disappears.
The distant instructions could not affect each other by using sequence characterization, which allows the model to identify the relationship between instructions more clearly.
Also, by using the linear interpolation algorithm, we can ensure the length of input sequences is the same.

\noindent \textbf{Padding Problem Avoidance.}
The padding problem could be solved because we only need to pad at both ends of the sequence before convolving.
What is more, compared to adding improper information to the image, the padding in the sequence will effectively mark the beginning and end of the corresponding program.
Therefore, the model could recognize the correct location of the instruction according to the padding.

\noindent \textbf{Semantic Loss Reduction.}
The semantic loss could be reduced because we input the whole binary instead of just taking a snippet.
By the linear interpolation algorithm, we can compress the semantic instead of ignoring it.
In this case, embedded viruses could also be included because all instructions in a program are input into the model.

Before scaling into the same length, we first normalize the whole sequences to make the value of elements between minus one and one stored in the float format. 
Because of the continuity of the real number field, the float format could represent more information than the integer format.
Therefore, this operation is necessary to reduce the semantic loss caused by the linear interpolation algorithm, as shown in Figure~\ref{fig:scale}.
By performing statistics on our datasets, we find that most files are around 256KB. 
Consequently, we scale all the input sequences to $2^{18}$ bytes.

In sequence format, all information among instructions will be correctly and better reserved.
The physical distance between two instructions reflects the true strength of the relationship.
This representation method also takes advantage of spatial locality in codes because the model will pay more attention to the instructions nearby instead of those far from each other with weak correlation.
Therefore, less interference will be received by the model when learning and detecting.

Another reason to use sequence characterization instead of binary-image converting is that executable files have a more pronounced before-and-after correlation than the planar correlation. 
That is why sequences could represent the programs better.


\subsection{Sequence Depthwise Separable Convolution}
\begin{figure}[]
\centering
\subfigure[]{
\includegraphics[scale=0.7]{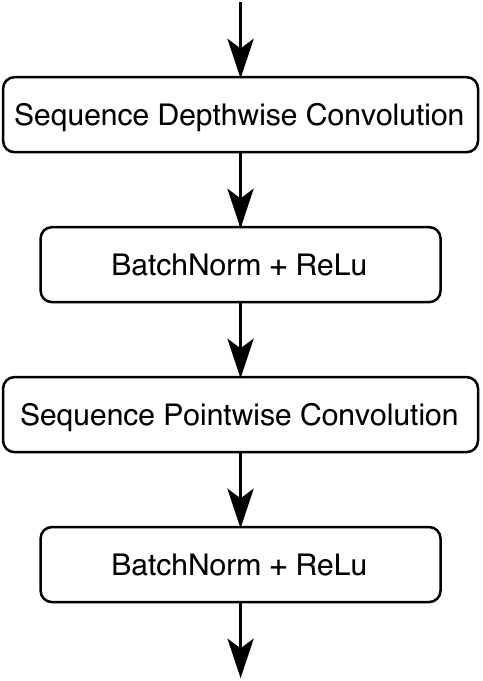}
\label{fig:SDSC:norm}
}
\subfigure[]{
\includegraphics[scale=0.7]{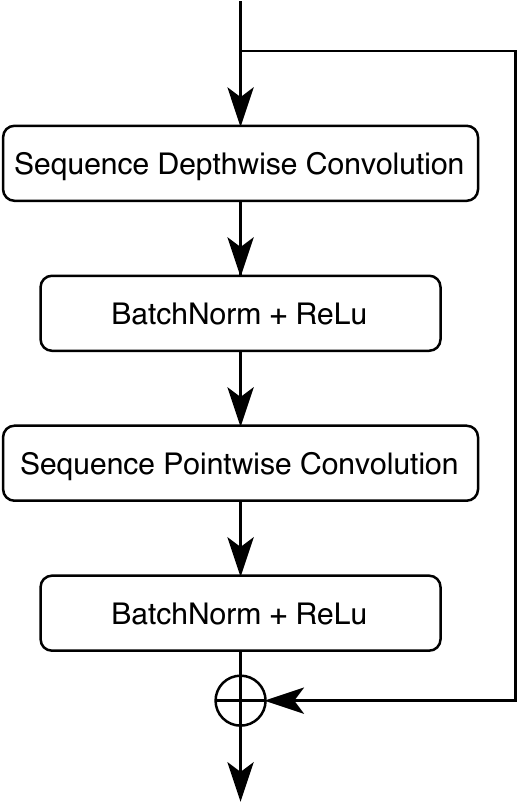}
\label{fig:SDSC:res}
}
\caption{Figure~(a) describes the standard SDSC block and Figure~(b) describes the residual SDSC block.}
\label{fig:SDSC}
\end{figure}

The sequence input format not only addresses the problem of semantic loss and contextual confusion but also compresses the scale of SeqNet.
The convolution kernels of SeqNet only need to extract features on a sequence that has just one dimension.
Compared to processing one dimension input, extracting features on two dimensions requires larger convolution kernels and more calculations. 
For example, as shown in Figure~\ref{fig:depthwise:our}, a $3 \times 3$ kernel used in images needs at least ten parameters~(including bias) while a $3 \times 1$ kernel used in sequences only needs at least four parameters~(including bias).

Based on sequence input and depthwise separable convolution~\cite{mobilenet}, we propose a method called Sequence Depthwise Separable Convolution~(SDSC) which needs fewer parameters and fewer calculations.
In SDSC, We use $3 \times 1$ kernels to replace the 2D depthwise convolution kernels in DSC.
By using the SDSC layers, SeqNet has a much smaller size than existing models.
In the following, we analyze the computation reduction compared to DSC.

Consider that the size of input is $n \times n \times c$ for the image format and $n^2 \times c$ for the sequence format where $c$ means the number of channels and $n$ means the width and the height. We also assume that the size of output is $n \times n \times c'$ and $n^2 \times c'$, the kernel size is $k \times k$ and $k \times 1$.

For common convolution, the number of calculations is
$$ Cal_{com} = n \cdot n \cdot c' \cdot c \cdot k \cdot k. $$
For DSC, the number of calculations is
$$ Cal_{DSC} = n \cdot n \cdot c \cdot k \cdot k + n \cdot n \cdot c' \cdot c. $$
For SDSC, the number of calculations is
$$ Cal_{SDSC} = n^2 \cdot c \cdot k + n^2 \cdot c' \cdot c. $$

The computation reduction is
$$ \frac{Cal_{DSC}}{Cal_{com}} = \frac{1}{c'} + \frac{1}{k^2}, $$
$$ \frac{Cal_{SDSC}}{Cal_{com}} = \frac{1}{c'k} + \frac{1}{k^2}. $$

Additionally, the input of SDSC is the one-dimensional data, so compared to the DSC, it is less susceptible to unrelated instructions.
In the experiments, we find that SDSC has great performance and successfully maintains the performance of SeqNet.

Based on SDSC, we use the following two main architectures of convolution blocks in SeqNet.
\begin{itemize}
\item \textbf{Standard SDSC Block.}
As shown in Figure~\ref{fig:SDSC:norm}, the standard SDSC block has three parts.
We use the batch normalization layers~\cite{BatchNormalization} to help the model learn the probability distribution of training samples better.
The ReLU~\cite{relu} activation function could accelerate the training process by making the model converge quickly.

\item \textbf{Residual SDSC Block.}
As shown in Figure~\ref{fig:SDSC:res}, the residual SDSC block combines the method used in ResNet~\cite{resnet}.
By skipping the SDSC block, we could effectively prevent the gradient from vanishing and build much deeper architecture.
\end{itemize}



\subsection{Model Architecture}
\begin{figure*}
\centering
\includegraphics[width=0.9\linewidth]{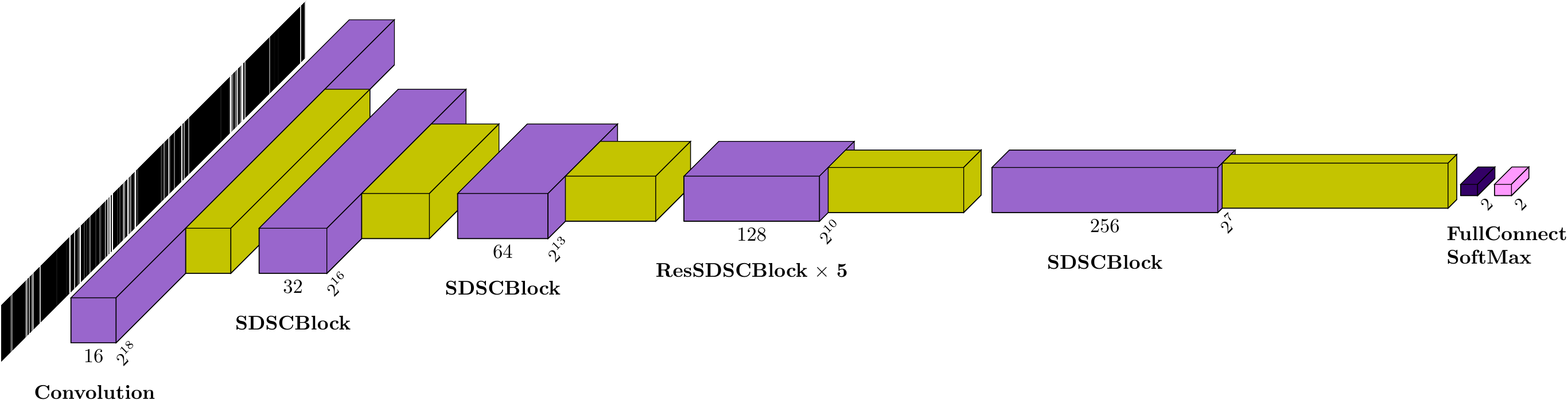}
\caption{Architecture summary of SeqNet.}
\label{fig:arch}
\end{figure*}
The construction of SeqNet is mainly based on SDSC, and Figure~\ref{fig:arch} explains the architecture. 
In order to reduce the number of parameters, we use smaller kernels and a deeper structure, which could also expand the receptive field.


The standard SDSC blocks are used to extract the features when downsampling the sequences.
For the first convolutional layer, we use a single common convolutional layer to embed the original input, and the size of the kernel is $3 \times 1$.
We set the size because the length of frequently used CPU instructions is usually three bytes.
For the high-level features, we use five residual SDSC blocks to analyze, which could also reserve the contextual and spatial information better compared to the fully connected layers. 
Besides, the residual SDSC block can make the model converge rapidly by preventing the gradient from vanishing.
The last two layers are a fully connected layer and a softmax layer.
The fully connected layer is used to classify the analysis result given by the front of the model.
The softmax layer could translate the result into the possibility format using the following formula.
$$P_i=\frac{e^{x_i}}{\sum_j e^{x_j}}, i=1,2$$
In this formula, $P$ means the translated result, and $x$ means the input of the softmax layer.
For the pooling layers, we use average pooling.

In our experiments, we find that when the number of residual SDSC blocks is five with 128 channels input and the number of fully connected layers is just one, the model performs the best.
SeqNet outputs the malicious possibility of a sample, and if the possibility is more than 50\%, the model will regard it as malware.
Depending on the output, for the loss function, we use the cross-entropy function.

In total, SeqNet has only about 136K parameters which are nearly one-tenth of MalConv, which we will discuss in Section~\ref{experiments}.


\section{Experiments}\label{experiments}
In this section, we will elaborate on our experiments.
All our experiments are conducted on a computer with a single GPU GeForce RTX 3090.

\subsection{Training Dataset}
Building a good training dataset is crucial for evaluating the performance of SeqNet.
Labeled correctly and having enough samples are necessary for showing the learning ability of SeqNet.

For the type of samples, we consider that PE malware is one of the main threats to electronic systems.
Also, there are many PE malicious samples, and it is easy to get enough PE samples.
Therefore, the following experiments are applied to a set of PE files because of their prevalence.

In this work, all the malicious samples come from VirusShare~\cite{vs}.
About 10,000 benign samples are provided by QI-ANXIN Corporation.
We also collect many benign samples from real personal computers to simulate the real environment in our daily lives.
To ensure that no virus is mixed in the benign samples, we detect all the files using VirusTotal~\cite{vt}.
If no AV engine regards it as malware in the VirusTotal report, we consider it a benign sample.

Operating system files and malware often have similar behaviors, which probably confuse the detection models and even a trained analyst~\cite{Survivalism}.
Therefore, in order to help SeqNet observe the general differences between malicious and benign programs and make SeqNet more robust, we add about 10,000 system files as benign data.
The system files are also checked by VirusTotal~\cite{vt} to ensure they are benign.

In total, we have obtained a training dataset of 72,329 binaries, with 37,501 malicious and 34,828 benign, and a validation dataset of 24,110 binaries, with 12,501 malicious and 11,609 benign. 
All the files in the datasets are PE files, and we remove the duplication by comparing SHA256 values.

We set the ratio of malicious and benign samples to about 1 to ensure the result is reliable.
For example, if the dataset only has malicious samples, the model might detect all malware by recognizing "4D 5A".
On the contrary, if we add enough benign samples into the dataset, the model could learn the real difference between malicious and benign samples.
        

\subsection{Measurement}
In our experiments, we measure SeqNet from two aspects, training cost and accuracy.

For the training cost measurement, we use the number of parameters to represent the scale of a model.
A larger model contains more neurons, which need more parameters to build.
Each parameter occupies constant memory during the training and predicting process.
Therefore, the number of parameters significantly influences the memory a model needs for training and predicting.
In order to measure the computational overhead for model inference accurately, we calculate the floating point of operations~(Flops) on each model by inputting a random binary.
We also measure the speed of a model by recording the time needed for an epoch, including the training and validation processes.

Besides accuracy, the measurements for performance we also use are precision, recall, and F1 score.
Their formulas are
$$Precision=\frac{TP}{TP + FP}$$
$$Recall=\frac{TP}{TP+FN}$$
$$F1 = 2 * \frac{Precision * Recall}{Precision + Recall}$$
where TP means the malicious samples which the model predicts correctly, FP means the benign samples which the model predicts incorrectly, and FN means the malicious samples which the model predicts incorrectly.

        
\subsection{Training Setup}
The model and training setup could significantly affect the training process and the result.
All models are trained for 70 epochs, and we select the validation results of the last 30 epochs to get the average accuracy and other metrics.
We set the batch size to 32 and choose Adam~\cite{Adam} as the optimizer of all models.
In order to ensure fairness in training, we apply the cross-entropy loss to all models.
    

\subsection{Model Evaluation}\label{experiments:models}
\begin{table*}
\centering
\begin{tabular}{cccccccc}
\hline
Model       & Parameters    & Accuracy           & Precision        & Recall  & F1 & MFlops & Speed \\
\hline
MobileNet   & 2.2M & 0.964$\pm$0.001 & 0.966$\pm$0.005 & 0.966$\pm$0.005 & 0.966$\pm$0.001 & 333 & 2min36s \\
MalConv     & 1.2M & 0.980$\pm$0.001 & 0.978$\pm$0.005 & 0.983$\pm$0.005 & 0.980$\pm$0.001 & 266 & 1h12min05s\\
MalConvGCT  & 1.2M & 0.980$\pm$0.001 & 0.978$\pm$0.004 & 0.983$\pm$0.005 & 0.981$\pm$0.001 & 1091 & 2h38min30s\\
SeqNet(Ours)& 136K & 0.974$\pm$0.002 & 0.974$\pm$0.008 & 0.975$\pm$0.007 & 0.974$\pm$0.002 & 193 & 2min51s\\
\hline
\end{tabular}
\caption{Comparison among models.
The metrics are the mean values and standard deviations of the last 30 epochs.
}
\label{tab:eval}
\end{table*}

\begin{figure}
\centering
\includegraphics[scale=0.5]{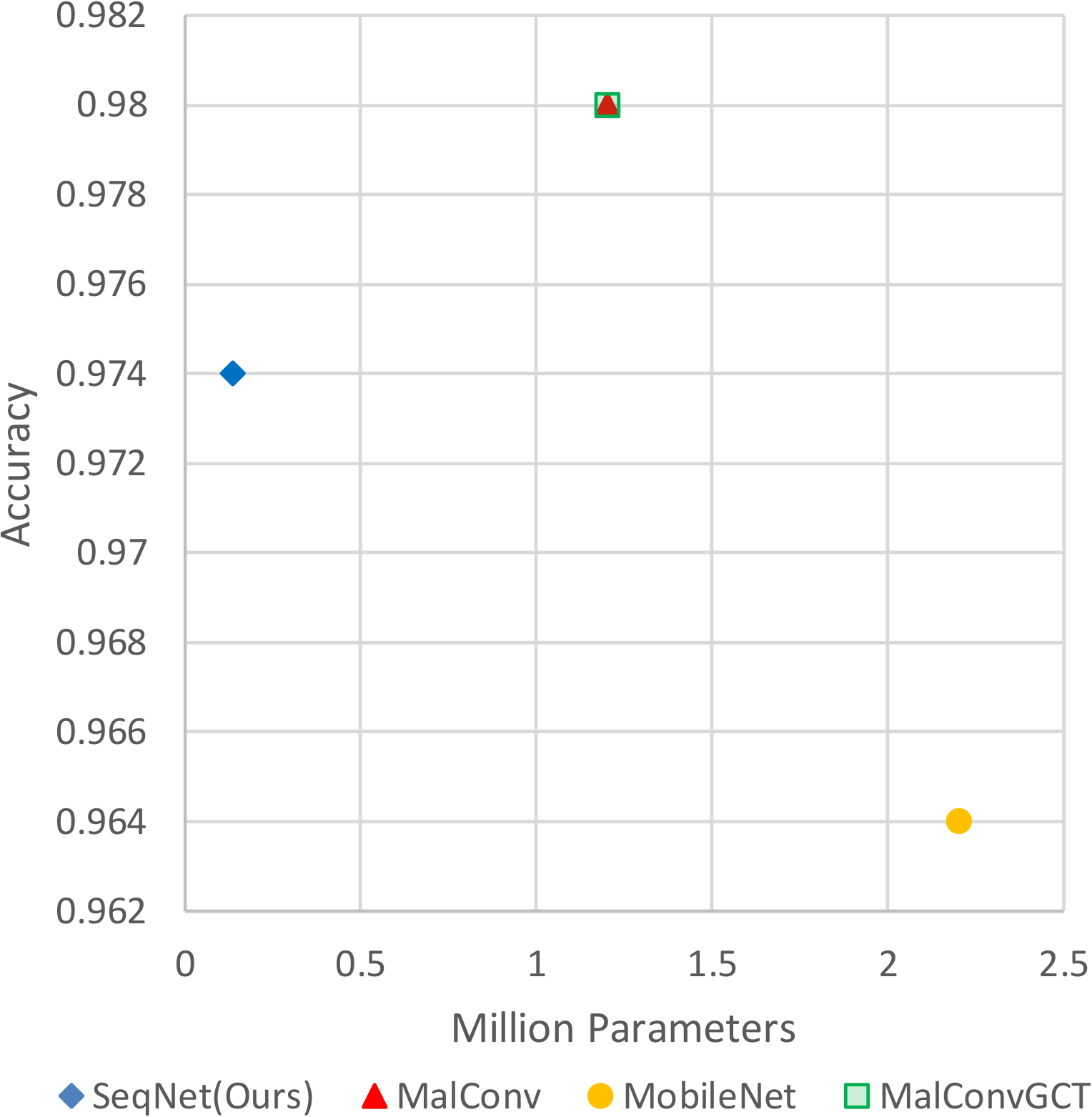}
\hspace{1em}
\caption{Accuracy vs Million Parameters}
\label{fig:accvspara}
\end{figure}

\begin{figure}
\centering
\includegraphics[scale=0.47]{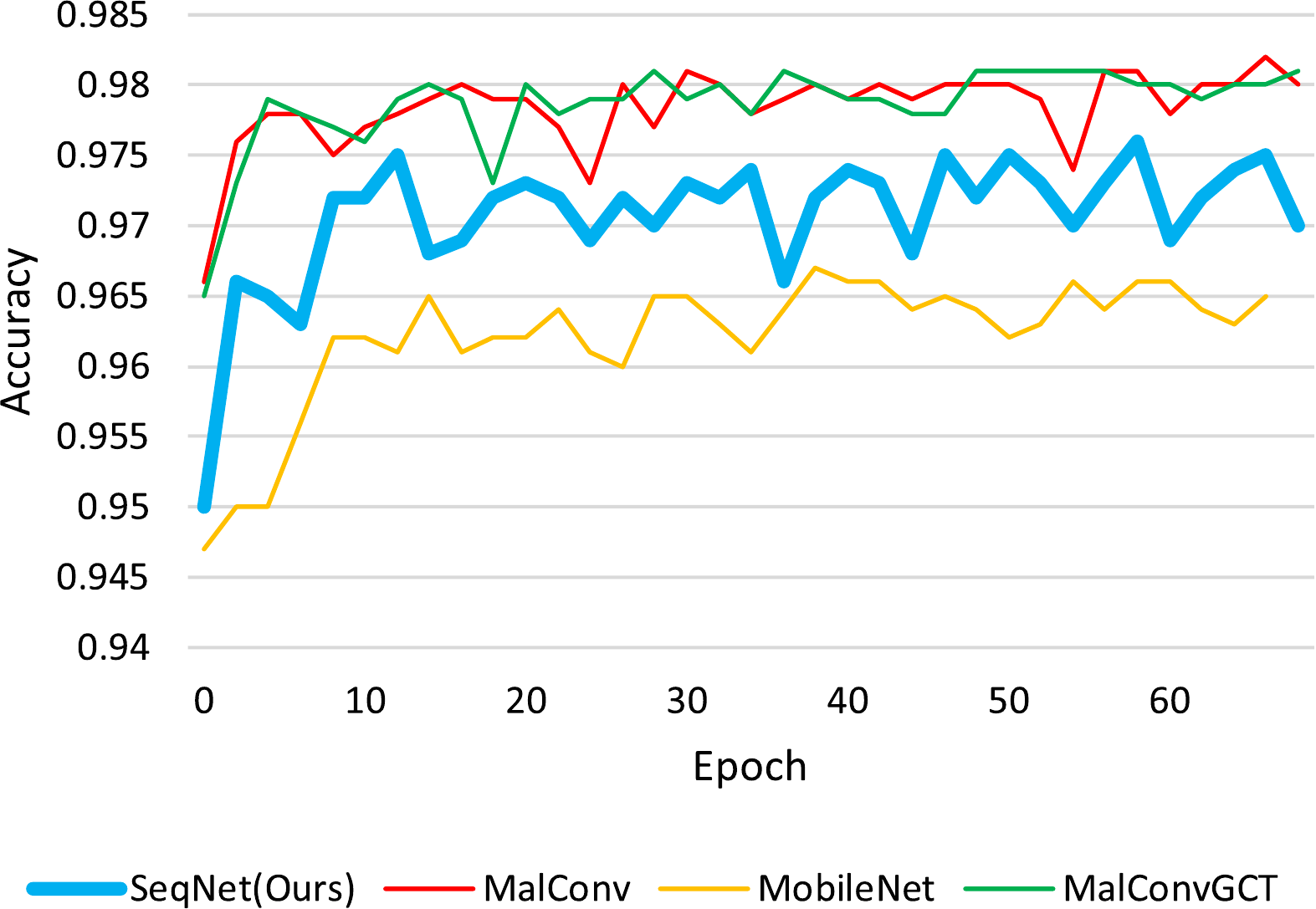}
\caption{Training processes among models.}
\label{fig:modelcmptraining}
\end{figure}
We choose several state-of-the-art binary-based methods as our baselines.
To reflect the total performance of image-converting-based models, we choose the well-known MobileNet~\cite{mobilenet} as the representative model.
We convert programs to RGB images as the input of MobileNet.
One byte maps one pixel in one channel during the converting process.

For file-cutting-based models, we select MalConv~\cite{malconv} and MalConvGCT~\cite{malconv2}.
Both ResNet and MobileNet are implemented by Pytorch~\cite{pytorch}, and we use ResNet18 as ResNet while MobileNetV2 as MobileNet.
We use the source codes provided by the author of MalConv and MalConvGCT and apply them to our experiments.
We add a softmax layer at the end of each model to translate the result into the possibility format when validating.

According to the number of parameters in Table~\ref{tab:eval}, we could find that SeqNet has the minimum parameters which are only about one-tenth of MalConv and MalConvGCT.
What is more, SeqNet maintains its performance in malware detection.
Figure~\ref{fig:accvspara} is the accuracy and model size comparison chart, and the position of a model in the upper left corner indicates that the model has a smaller size and a higher accuracy.
The precision also shows that SeqNet has a low possibility of misunderstanding benign samples.
The recall implies that SeqNet might have the ability to prevent evasion.

In the training process, We find that most models achieve 90\% accuracy after the first epoch.
During our training, we find that SeqNet only needs about two and a half minutes to complete an epoch compared to MalConv which is about an hour.
Accordingly, we can see that the tiny size of SeqNet leads to low computational overhead, and the convolution-based architecture accelerates the training and inference.
        

\subsection{Further Evaluation \& Ablation Study}
\begin{table*}
\centering
\begin{tabular}{cccccc}
\hline
Model       & Parameters    & Accuracy       & Precision     & Recall & F1 \\
\hline
SeqNet2D    & 141K & 0.964$\pm$0.001 & 0.965$\pm$0.004 & 0.965$\pm$0.004 & 0.965$\pm$0.001 \\
SeqNetConv  & 381K & 0.975$\pm$0.003 & 0.974$\pm$0.008 & 0.977$\pm$0.008 & 0.976$\pm$0.003 \\
SeqNet2DConv& 1.1M & 0.967$\pm$0.001 & 0.969$\pm$0.004 & 0.967$\pm$0.006 & 0.968$\pm$0.001 \\
SeqNetDeep  & 270K & 0.976$\pm$0.001 & 0.976$\pm$0.005 & 0.978$\pm$0.005 & 0.977$\pm$0.001 \\
SeqNetShal  & 101K & 0.970$\pm$0.002 & 0.972$\pm$0.007 & 0.971$\pm$0.009 & 0.972$\pm$0.002 \\
SeqNet(Ours)& 136K & 0.974$\pm$0.002 & 0.974$\pm$0.008 & 0.975$\pm$0.007 & 0.974$\pm$0.002 \\
\hline
\end{tabular}
\caption{Analysis and ablation study results of SeqNet.}
\label{tab:analysis}
\end{table*}

Next, we will describe further experiments to discuss contextual confusion avoidance, model shrinkage, and architecture design in SeqNet.
We apply several simple changes to SeqNet to test our assumptions.

\noindent \textbf{Contextual Confusion.}
To prove that the contextual confusion exists, we change the SeqNet into SeqNet2D which could input $512\times512$ images.
We set the image size to $512 \times 512$ because it could the same amount of information as the sequence.
The architecture of SeqNet2D which uses depthwise separable convolution layers is highly similar to SeqNet.
Table~\ref{tab:analysis} makes sure that the image converting method causes confusion that could confuse the network when learning features from programs.
MobileNet uses images as the input, so Table~\ref{tab:eval} also reflects the existence of contextual confusion through the comparison among SeqNet and ResNet.

\noindent \textbf{Model Shrinkage.}
SDSC layers reduce the amount of the parameters by both decreasing input dimension and factorizing convolution.
We replace all SDSC layers with common convolutional layers in SeqNet and name the new model SeqNetConv to reflect the effect of factorizing convolution.
To validate the role of dimension decreasing in model shrinkage, we use SeqNet2D as the experimental subject.
We also apply common convolutional layers to SeqNet2D and call the new model SeqNet2DConv to show the reduction effect of the two methods.
As shown in Table~\ref{tab:analysis}, we can see that the dimension effectively decreasing reduces the number of parameters, and the convolution factorizing reduces further.
Through the result, it is obvious that the SDSC layers effectively shrink the model with performance maintained.
We also find that the fewer parameters make the model quicker to converge when training.

\noindent \textbf{Architecture Design.}
During our experiments, we also find that deeper structures may not perform better.
We adjust the depth of SeqNet.
The model with the deeper architecture is called SeqNetDeep, and the more shallow architecture is called SeqNetShal.
Similar to our intuition, the shallower architecture significantly makes the model perform worse.
However, the result in Table~\ref{tab:analysis} shows that the deeper architecture cannot improve the performance apparently but enlarge the size of the model.
We assume that this phenomenon is probably because of the simple mapping relationships from the binary to the possibility of malware, which does not need a complex neural network to fit. 
Another possible reason is that the deeper structure makes the network hard to be trained, which might lead to lower accuracy.

Another phenomenon we find is that the dilated convolution could not improve the performance effectively.
We consider that it is because compared to the length of the input, using or not using the dilated convolution does not have much effect on the receptive field of the model.
    
\subsection{Robustness Evaluation}

\begin{figure}
\centering

\includegraphics[scale=0.35]{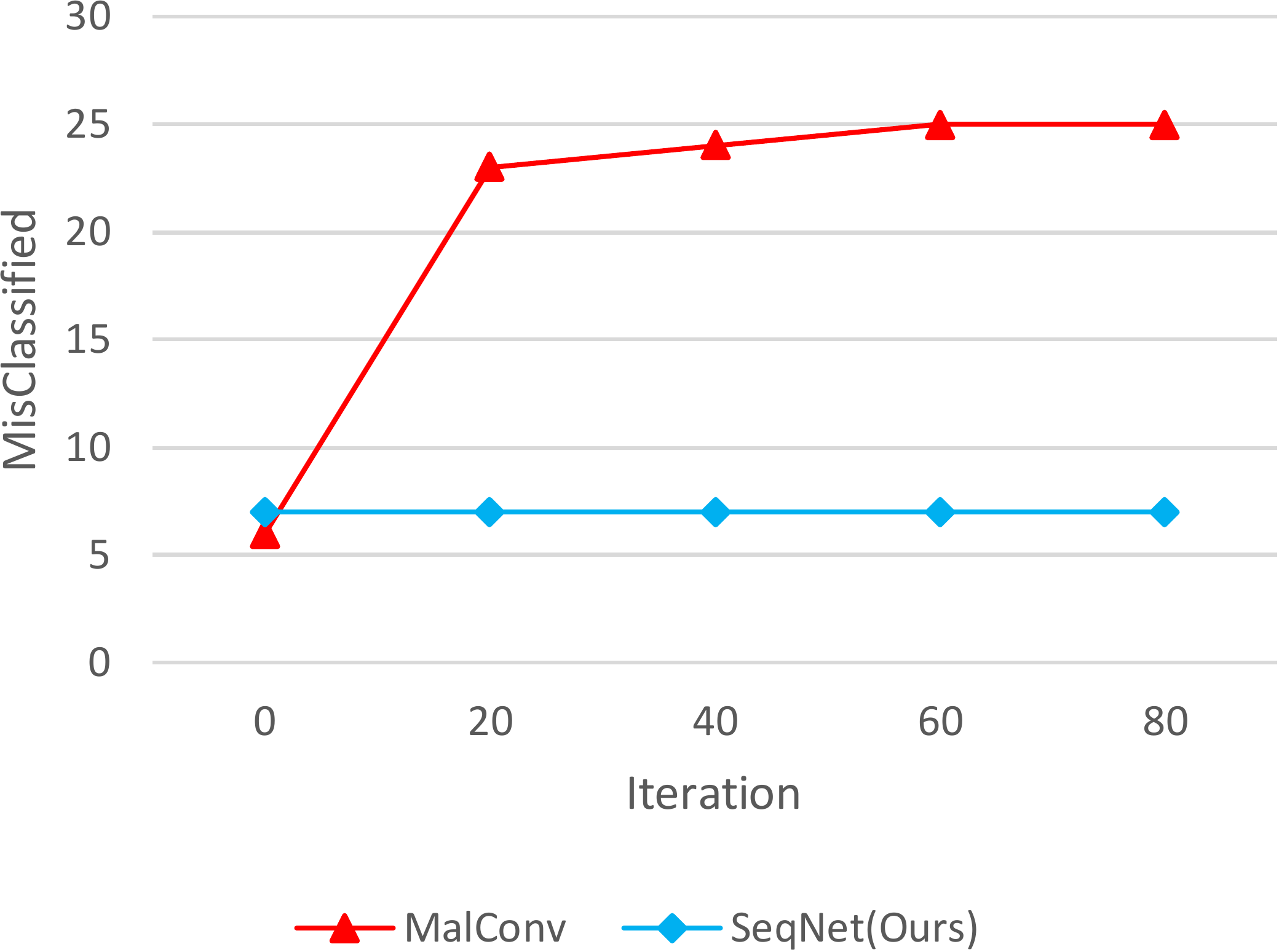}
\hspace{2em}
\caption{Robustness Evaluation.}
\label{fig:robustness}
\end{figure}

We also check the robustness of SeqNet and compare it with MalConv.
There has abundant research about attacking deep models~\cite{honeypots, pgd, got-a-model, badencoder, you-see-what-i-want, composite-backdoor, Wang_2021_CVPR}.
However, different from traditional attacks against neural networks on image-related tasks, we could not straightforwardly add perturbation on binaries because it might make binaries not executable.
Also, it is hard for us to adapt attack strategy based on extracted features~\cite{explanation-attack} to raw-binary models.
Therefore, for attack strategy, we apply the approach in~\cite{againstmalconv} which injects a short poison binary in the padding part at the end of the input.

Because of the different input formats between MalConv and SeqNet, we equivalently adapt the poison binary generating approach.
Compared to selecting the closest embedding vector along the gradient, the poison generation process of SeqNet steps as the following formula.
$$
x_i^{t+1} = x_i^t - \frac{1}{256}sign(\frac{\partial y^t}{\partial x_i^t})
$$
$$
sign(x) = \left\{
\begin{array}{cc}
1, &x > 0 \\
0, &x = 0\\
-1, &x < 0
\end{array}
\right.
$$
$x_i^t, y^t$ denote an element of the normalized poison and the prediction given by SeqNet in the $t$-th iteration, respectively.

In order to make the attack strategy effective under the limited binary length, we randomly select 500 available samples from the validation malicious dataset.
We set the length of poison to 32000 bytes, which the whole binaries for MalConv are fixed to 16000000 bytes, and we progressively increase the poison generation iterations.
We test the number of samples misclassified by SeqNet and MalConv.

The results are displayed in Figure~\ref{fig:robustness}.
We see that SeqNet has a great defensive capability against the poison binary attack.
We assume that this phenomenon is because of the vulnerability caused by the padding part in the file-cutting approach.
File-cutting-based models often see incomplete binary while training.
Therefore, the padding in file-cutting gives attackers the chance to confuse models, while the models are not sure whether the poison binary is one of the parts of the sample.
Compared to file-cutting, our method could alleviate this problem by inputting the whole binary.
However, we consider that this theory still needs to be further verified, and we might research it in future work.

\subsection{Case Study}
\begin{figure*}
\centering
\includegraphics[scale=0.035]{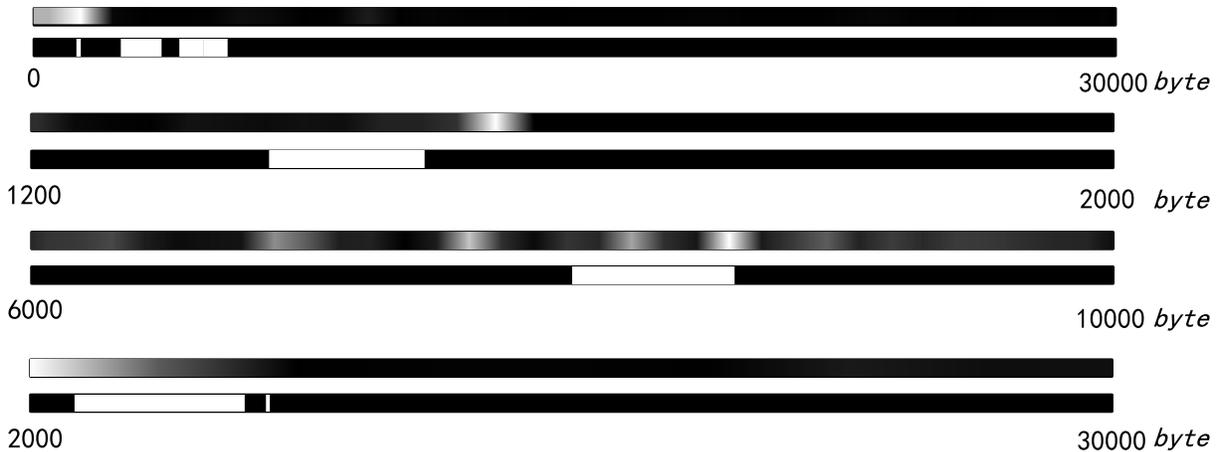}
\caption{The comparison between manual location maps and SeqNet heatmaps of four samples.
In each pair, the upper is the heatmap generated by Grad-CAM~\cite{grad-cam} and the lower is the malicious code location map given by analysts.
In heatmaps, the whiter parts are more critical to the prediction given by SeqNet.
Similarly, the white parts in manual location maps are the locations of the malicious codes.
The scale under each pair is the file offset of the corresponding two maps.
}
\label{fig:explain}
\end{figure*}
To better understand what SeqNet has learned, we randomly select four samples and use the Grad-CAM~\cite{grad-cam} explanation technique to generate heatmaps so that we could visualize which part affects the result most.
Also, we manually analyze the corresponding samples to verify whether SeqNet finds the right malicious codes.
In manual analysis, we disassemble the samples by IDA Pro~\cite{IDA} and precisely locate the malicious functions or codes.

In order to better plot the results, we extract the key part of the heatmaps and apply the following normalization formula to the snippets.
$$
x_i = \frac{x_i}{max(X)}, x_i \in X
$$
where $X$ denotes the snippet.
The activation maps used for heatmaps are generated by the last convolution and ReLU layer of SeqNet, because of the remained spatial information encoded by convolutional layers.
We also mark the manual location results on raw binaries for better comparison.

Figure~\ref{fig:explain} implies the comparison between manual location and Grad-CAM-based explanation.
We see that the local activated positions are closed to the malicious parts located by analysts, which reflects that SeqNet might find the malicious codes and give its reliable detection.

During our explanation, we find that there are many noises in the whole heatmaps.
We consider that this might be because of the possible latent abnormal statistics~\cite{overinterpretation} and a few wrong labels~\cite{AndroidClassificationServices} in our dataset.
However, it is a pity that we find it hard to collect more benign samples due to the neglect of benign files in the academic community.
We hope that we could explore this phenomenon in future work.

Also, we find that the PE headers often affect SeqNet greatly.
This might imply that the PE header contains malicious information in malware.
More details could be found in Appendix.

\section{Discussion}\label{discussion}
In this section, we will take about the limitation of our works and suggest some future works for further study.
\subsection{Limitation}
Although SeqNet performs well, there are still several limitations in our works.

\noindent \textbf{Remained Semantic Loss.}
Although we effectively reduce the semantic loss, the input of SeqNet still cannot contain all semantics.
If the sequence is too long, the sequence will be compressed during the interpolation process, and the compressed sequence is not able to represent all original information.
Also, if the sequence is too short, the sequence will be expanded, which might confuse SeqNet.
The architecture of SeqNet determines that the input must have the same size, which is a limitation of SeqNet.

\noindent \textbf{Lack of Benign Samples.}
The main difficulty we face is the lack of benign samples.
We can get plenty of malware collection websites, but authoritative benign sample providers are hard to find.
In order to sample evenly, it is improper to expand the training dataset only by adding malicious samples, which might reduce the performance of SeqNet and make the experiment results unreliable.
As a result, it is difficult to train neural networks on a much larger dataset with enough benign samples.


\noindent \textbf{The Quality of Labels.}
Besides the lack of benign samples, the quality of labels might be a potential problem.
Due to the few authoritative providers, we cannot guarantee that all the benign samples in our training and validation datasets are labeled correctly.
All malicious samples in our datasets are collected from VirusShare without manual reconfirmation.
Several papers have inspected the quality of malware labels and found it probably cannot reach what we expect~\cite{AndroidClassificationServices}.
Although these limitations might have a few impacts on the training process of SeqNet, we assume that the several incorrect labeled samples could not significantly affect the overall performance.

\noindent \textbf{Possible Vulnerabilities.}
Adversarial attacks are the risk of most neural networks, and ours is no exception.
A motivated adversary could pollute the training dataset and evade the detection of SeqNet.
Also, attacking based on the gradient is an effective way to confuse deep learning models~\cite{pgd, againstmalconv, AdversarialPerturbations}.
On the contrary, there are also plenty of solutions to this problem~\cite{six_principles, DL-FHMC, RobustPDF, Differential, NAS-robust}.
Although SeqNet could defend against several attacks, we still cannot completely guarantee the safety of SeqNet.
Also, the robustness principle of SeqNet needs us to explore further.

\subsection{Future Work}
In this work, we propose an efficient automatic malware detection neural network called SeqNet.
SeqNet mainly aims at automatic and efficient detection and could be quickly trained with low training costs on raw binaries.
Nevertheless, many works still need to be done in the future. 

One of the biggest obstacles to malware detection research based on deep learning is the lack of industrial-sized publicly available datasets.
The researchers require authoritative credible datasets that contain not only malicious features but also raw binary sequences.
We will build a larger dataset to further evaluate the performance of SeqNet.
Also, enough benign samples are necessary for further study.
We suppose that when using deep learning models for detection, malware analysis should not only focus on malicious samples but also on benign samples.

Because neural networks are black-box models, the reliability of malware detection neural networks might be suspicious.
Although SeqNet gives us great results, we still cannot fully explain the reason.
Therefore, using deep learning algorithms to detect malware in practice still needs further research.
Through our experiments, we suppose that neural networks might have great potential capacities for malware detection, and we are looking forward to the big breakthrough neural networks make in this field.

The robustness of SeqNet still needs further research.
We still lack experiments and studies in this area.
In future work, we will explore the robustness of the model deeper and make more efforts to improve and analyze it.
        
\section{Related Work}\label{related work}

\noindent\textbf{Feature-based Models.} 
In the early works, deep learning models are trained from carefully crafted malware features.
Yuan et al. used deep learning algorithms to detect Android malware based on 202 manual features extracted through static and dynamic analysis~\cite{Droid-Sec}.
Saxe et al. designed a network with four layers and trained the model with the static features extracted from PE files~\cite{deepflow}.
Huang et al. used 4.5 million samples as the training dataset and used the random projection to reduce feature space from 50000 dimensions to 4000 dimensions~\cite{mtnet}.
Zhang et al. combined dynamic analysis techniques with deep learning algorithms and used the API calls with arguments to train their network~\cite{dynamic}.
Xu et al. devised a two-stage inference framework to detect Android malware on the extracted information~\cite{deeprefiner}.
Li et al. leveraged Autoencoder~(AE) to find the mutation of malicious features and got great performance~\cite{robust-android}.
Han et al. extracted installation graphs from viruses and applied Graph Neural Network~(GNN) to malware detection~\cite{sigl}.

\noindent\textbf{Binary-based Models.} 
Nowadays, automatic feature extraction is one of the trends of neural networks with less human intervention and better performance than traditional feature engineering.
Raff et al. devised an architecture called MalConv which could learn directly from the raw PE binary samples without manual feature selection~\cite{malconv, malconv2}.
Krc{\'{a}}l et al. designed a simple CNN which learns from PE raw byte sequence without domain-specific feature selection, and this work achieved a high AUC score, especially on the small PE malware samples~\cite{M2018Deep}.

\section{Conclusion}\label{conclusion}
In this paper, we introduce SeqNet, an efficient automatic malware detection model based on deep learning.
Compared to existing models, SeqNet has a much smaller size with enough detection accuracy, which can be more quickly trained on raw binaries and probably has more application scenarios.
Additionally, the training and detection process of SeqNet needs little human intervention.

We hope this model will inspire more researchers to develop better architectures for more applications, not only malware detection.
Also, we expect that deep learning algorithms will be widely used in practice for malware detection.

\section*{Availability}
We make the codes and data of SeqNet available to the research community to promote the adoption of SeqNet in security research and deployment. The SeqNet project website is at
\url{https://github.com/Darren-8/SeqNet.git}.

\section*{Acknowledgements}\label{acknowledgements}
This research has been partially supported by the National Natural Science Foundation of China (61872202), the Natural Science Foundation of Tianjin (19JCYBJC15500), 2019 Tianjin New Generation AI Technology Key Project under Grants (19ZXZNGX00090), Tianjin Key Research and Development plan (20YFZCGX00680). 
We also thank the authors of VirusShare~\cite{vs}, for their public dataset used in our evaluation, and VirusTotal~\cite{vt} for providing us the malware scan results.

{\footnotesize \bibliographystyle{acm}
\bibliography{reference}}

\appendix

\section*{A.\  Training Details}
\begin{figure*}[htb]
\centering
\subfigure[]{
\includegraphics[scale=0.6]{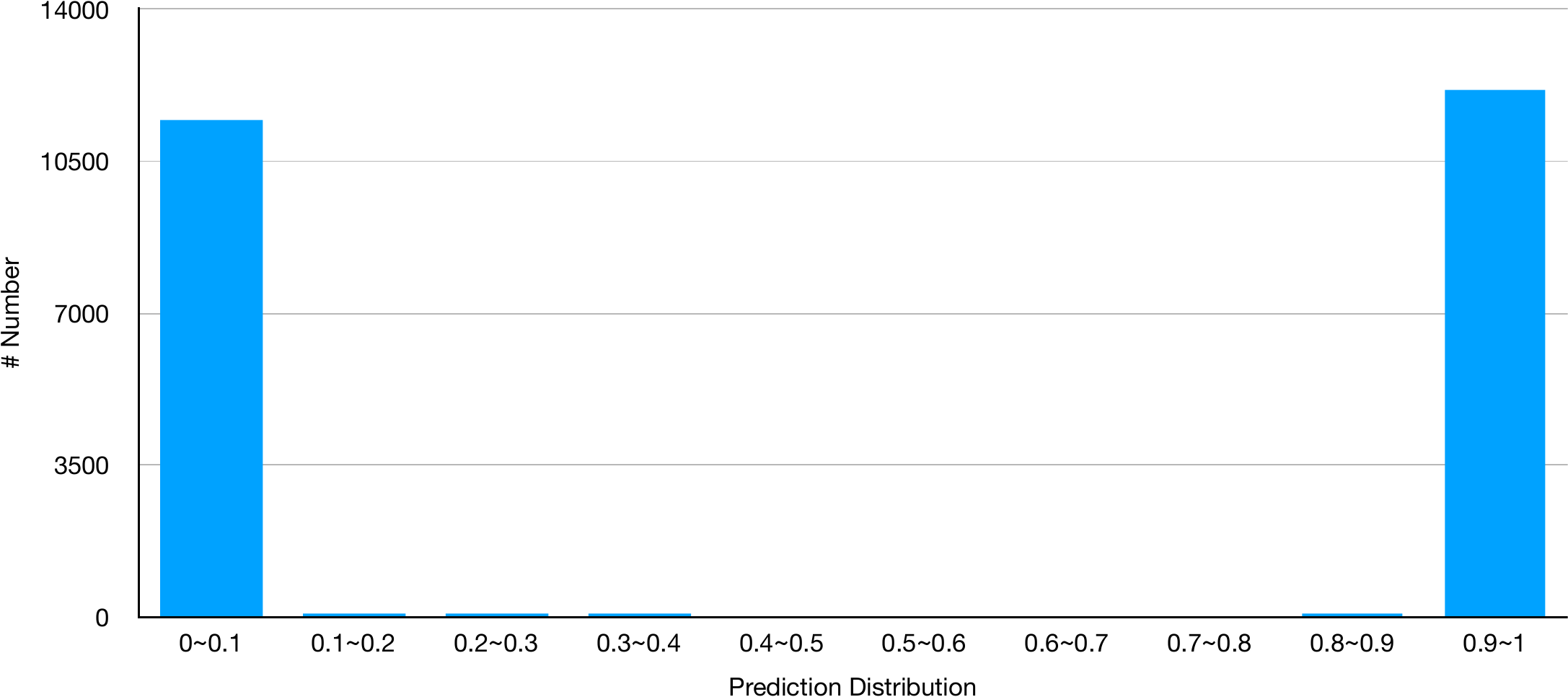}
\hspace{4em}
}
\subfigure[]{
\includegraphics[scale=0.6]{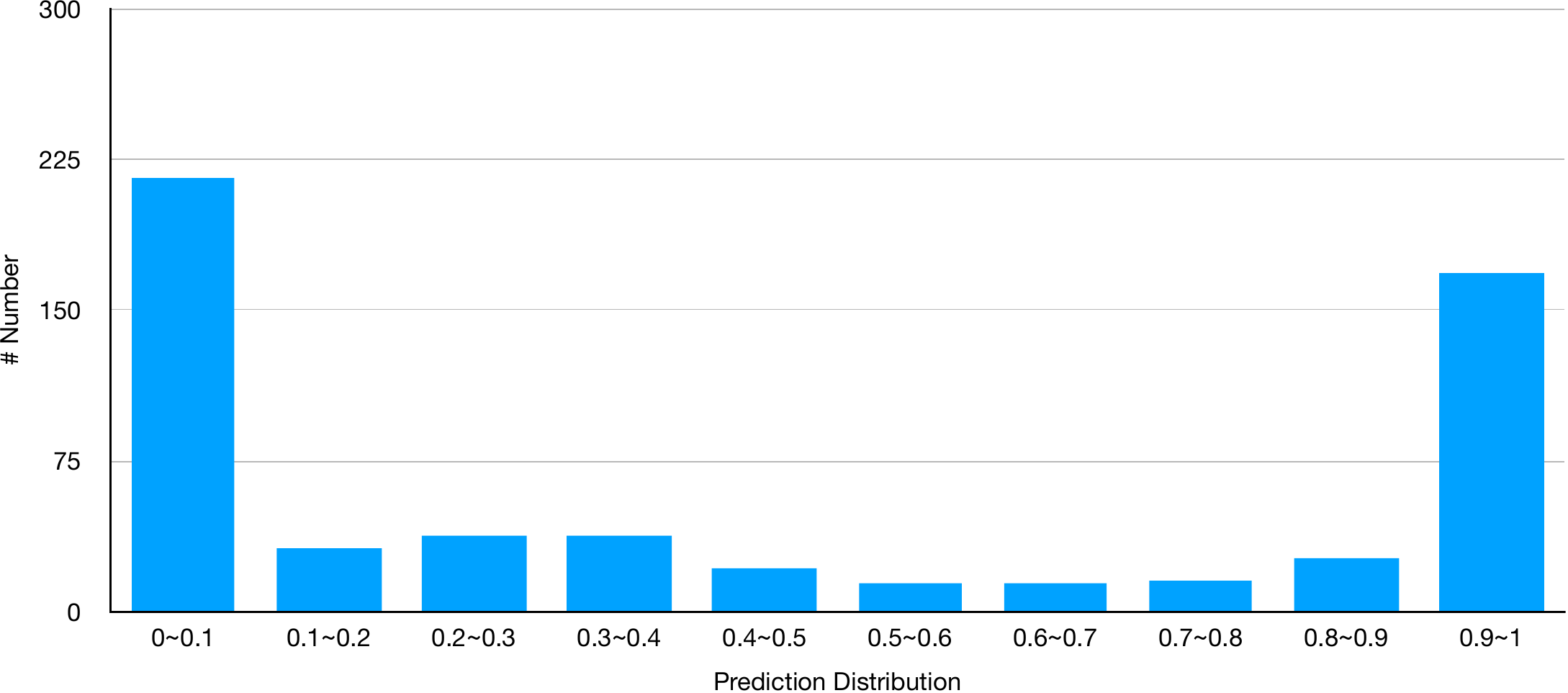}
\hspace{3em}
}
\caption{The prediction result distribution of SeqNet on the whole validation dataset.
(a) is the distribution of all the prediction results.
(b) is the distribution of the misclassified predictions.
}
\label{fig:appendix_pred_dis}
\end{figure*}

\begin{figure*}
\centering
\includegraphics[scale=0.6]{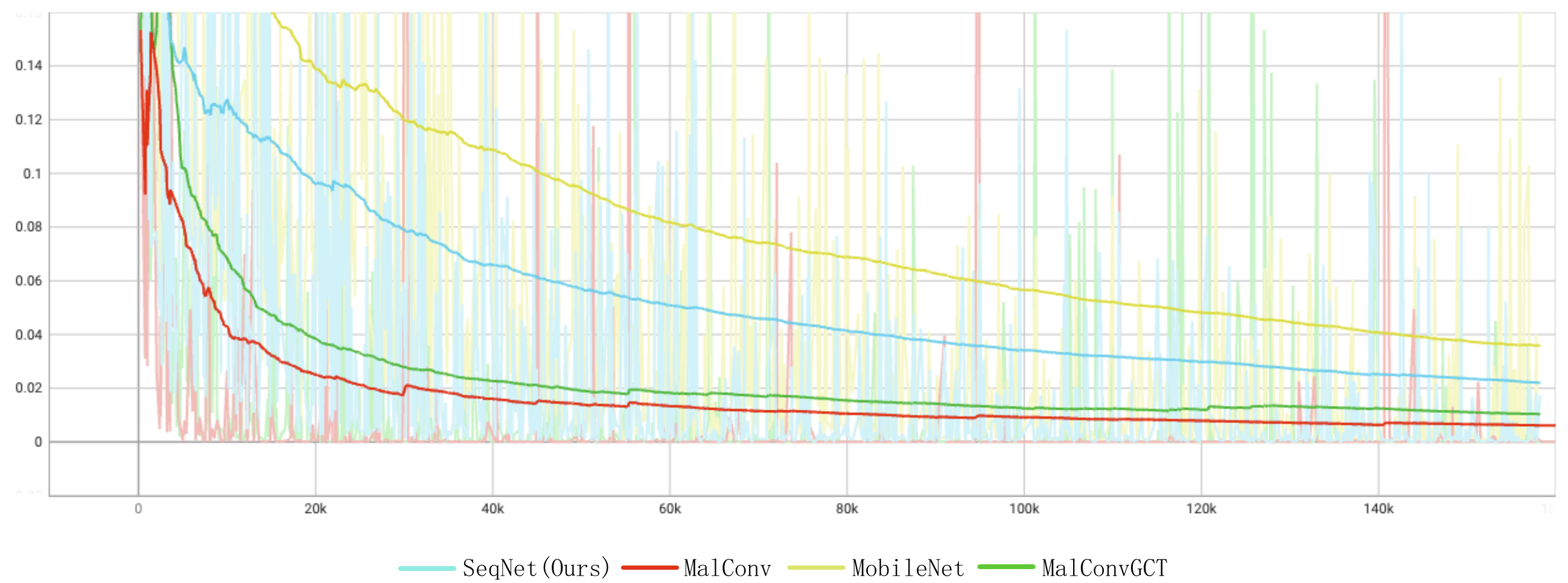}
\caption{Training processes among models in model evaluation.
The $X$ axis and $Y$ axis denote the training loss and training steps, respectively.
}
\label{fig:appendix_train}
\end{figure*}

\begin{figure*}
\centering






\includegraphics[scale=0.6]{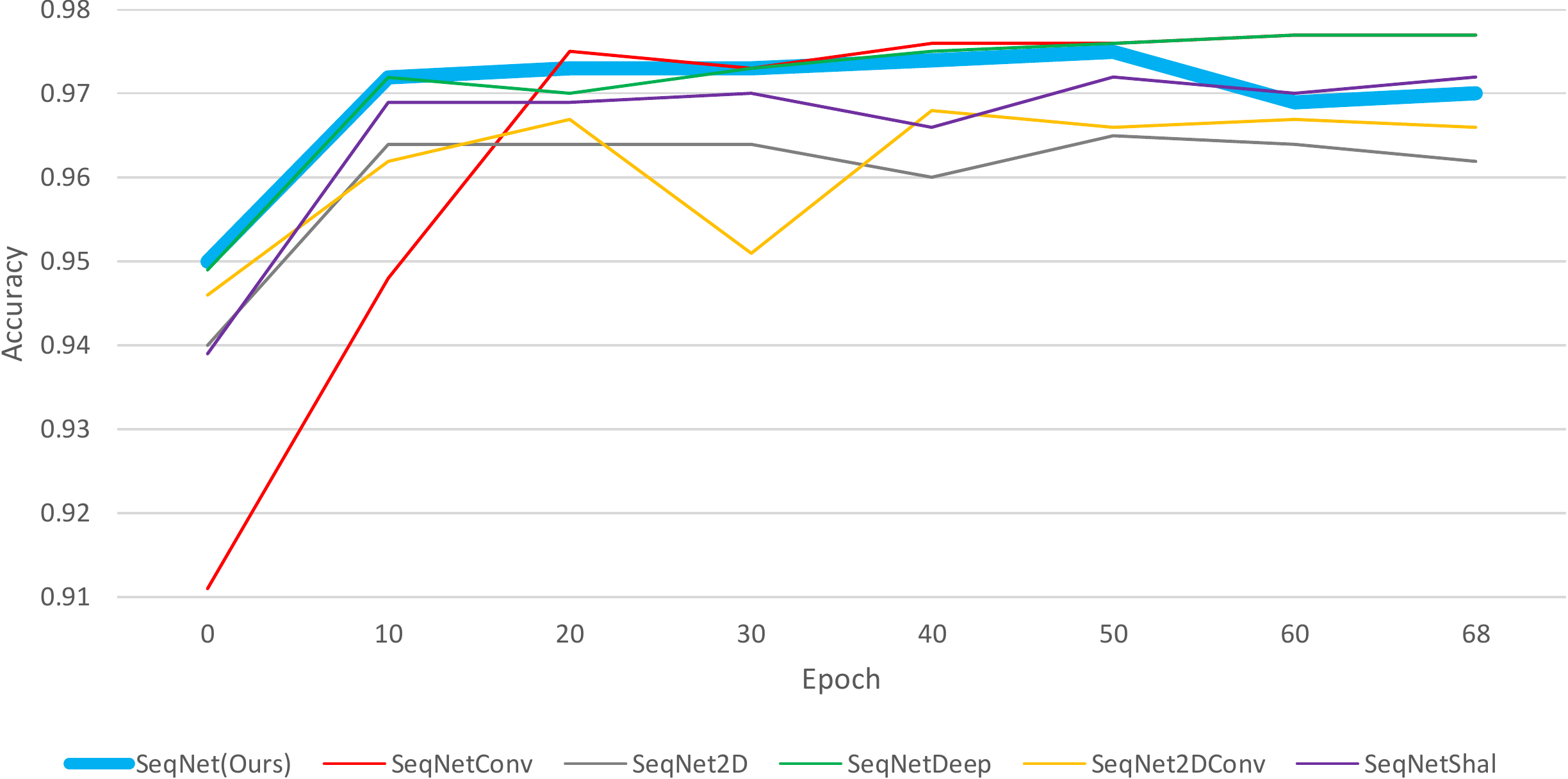}
\caption{Training processes among models in analysis and ablation study.}
\label{fig:appendix_ablationstudyprocess}
\end{figure*}
To further explore the performance of SeqNet, we record the training processes and the prediction result distribution on the validation dataset in the last epoch.
The results are plotted in Figure~\ref{fig:appendix_train} and Figure~\ref{fig:appendix_pred_dis}.
We see SeqNet has strong confidence on most samples, and all models converge quickly.

We also record the accuracy changes on the validation dataset in our ablation study, as shown in Figure~\ref{fig:appendix_ablationstudyprocess}.

\section*{B.\  Holistic Heatmap Analysis}
\begin{figure*}
\centering
\subfigure[]{
\includegraphics[scale=0.06]{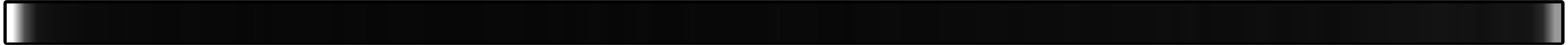}
}
\subfigure[]{
\includegraphics[scale=0.06]{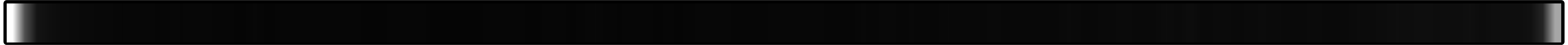}
}
\subfigure[]{
\includegraphics[scale=0.06]{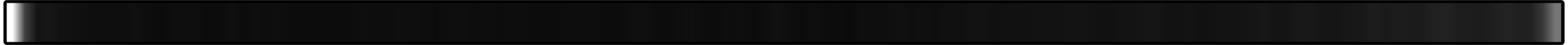}
}
\caption{The average heatmaps on the whole validation dataset.
(a) is the average heatmap on all samples.
(b) is the average heatmap on malicious samples.
(c) is the average heatmap on benign samples.
}
\label{fig:appendix_explain_avg}
\end{figure*}

\begin{figure*}
\centering
\subfigure[]{
\includegraphics[scale=0.06]{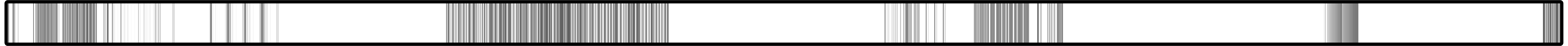}
}
\subfigure[]{
\includegraphics[scale=0.06]{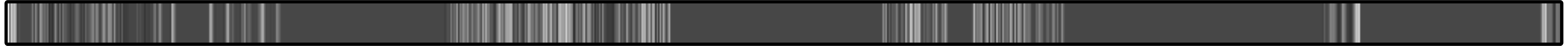}
}
\subfigure[]{
\includegraphics[scale=0.06]{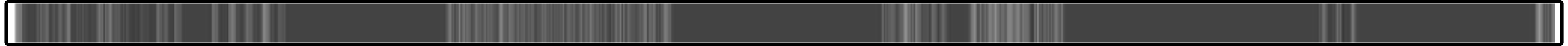}
}
\subfigure[]{
\includegraphics[scale=0.06]{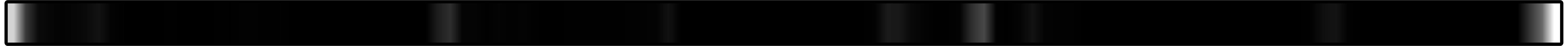}
}
\caption{The heatmaps on the different layers of SeqNet.
(a) is the heatmap after the first convolutional layer.
(b) and (c) are the heatmaps during downsampling.
(d) is the heatmap after all convolutional layers.
}
\label{fig:appendix_explain_arch}
\end{figure*}
We apply Grad-CAM~\cite{grad-cam} on the whole validation dataset and get all the heatmaps of the last convolution layer.
Then we sum all the heatmaps and get the average heatmap to show the general result.
Figure~\ref{fig:appendix_explain_avg} reflects that the activated features are often at the beginning and the end of the whole binaries, especially the beginning.
We consider that this phenomenon is because the PE header usually aggregates the features of the whole file.

To understand how SeqNet analyzes the binaries, we randomly select a sample and get the heatmaps of different layers by using Grad-CAM, shown in Figure~\ref{fig:appendix_explain_arch}.
We can see that the first convolutional layer extracts abundant semantics from the whole binaries.
During downsampling, the heatmap implies that features start to aggregate.
After processed by the last convolutional layer, the activated parts become sparse.
At last, the fully connected layer will give its result based on these features.

This result is similar to~\cite{deconv}.
The front layers often extract underlying features, and the layers at the end tend to analyze the overall information.
\end{document}